# High critical current density and improved flux pinning in bulk $MgB_2$ synthesized by Ag addition

Chandra Shekhar [a)], Rajiv Giri, R. S. Tiwari and O. N. Srivastava

*Department of Physics Banaras Hindu University, Varanasi-221005, India*

S. K. Malik

*Tata Institute of Fundamental Rresearch, Mumbai-400005, India*

**ABSTRACT**

In the present investigation, we report a systematic study of Ag admixing in $MgB_2$ prepared by solid-state reaction at ambient pressure. All the samples in the present investigation have been subjected to structural/ microstructural characterization employing x-rays diffraction (XRD) and transmission electron microscopic (TEM) techniques. The magnetization measurements were performed by physical property measurement system (PPMS). The TEM investigations reveal the formation of MgAg nanoparticles in Ag admixed samples. These nanoparticles may enhance critical current density due to their size (~ 5-20 nm) compatible with coherence length of $MgB_2$ (~ 5-6 nm) . In order to study the flux pinning effect of Ag admixing in $MgB_2$, the evaluation of intragrain critical current density ($J_c$) has been carried out through magnetic measurements on the fine powdered version of the as synthesized samples. The optimum result on intragrain $J_c$ is obtained for 10 at.% Ag admixed sample at 5K. This corresponds to ~$9.23 \times 10^7$ $A/cm^2$ in self-field, ~ $5.82 \times 10^7$ $A/cm^2$ at 1T, ~ $4.24 \times 10^6$ $A/cm^2$ at 3.6T and ~ $1.52 \times 10^5$ $A/cm^2$ at 5T. However, intragrain $J_c$ values for $MgB_2$ sample without Ag admixing are ~ $2.59 \times 10^6$ $A/cm^2$, ~ $1.09 \times 10^6 A/cm^2$, ~ $4.53 \times 10^4$ $A/cm^2$ and $2.91 \times 10^3$ $A/cm^2$ at 5 K in self field, 1T, 3.6T and 5T respectively.. The high value of intragrain $J_c$ for Ag admixed $MgB_2$ superconductor has been attributed to the inclusion of MgAg nanoparticles into the crystal matrix of $MgB_2$, which are capable of providing effective flux pinning centres. A feasible correlation between microstructural features and superconducting properties has been put forward.

----------------------------------------------------------

[a)]Electronic mail: chand_bhu@yahoo.com

**INTRODUCTION**

The recent discovery of superconductivity in intermetallic compound $MgB_2$ has generated tremendous interest because of its potential for applications at high magnetic field [1,2]. The critical current density ($J_c$) and upper critical field ($H_{c2}$) are two most important parameters of any superconductor for practical applications. $MgB_2$ possesses higher upper critical field than that of conventional superconductors (NbTi, $Nb_3Sn$ etc.). It appears to be the first promising intermetallic superconductor for applications in temperature range 20-30 K at high magnetic field [3,4]. Another important feature of $MgB_2$ is that now it is believed that unlike high temperature cuprate superconductors $MgB_2$ does not contain intrinsic obstacles to current flow between the grains [5]. Evidence for strongly coupled grains have been found even for randomly aligned, porous and impure samples [6,7], thus providing high feasibility of scaling up the material to form shapes like wires and tapes [8-10]. Large engineering applications have been hampered so far by low density and poor flux pinning behaviour of $MgB_2$ which induces degradation of $J_c$ in high magnetic fields. Many researchers have attempted to improve the flux pinning behaviour through several types of processes such as high energy ion-irradiation [11], chemical doping using different metallic and non-metallic phases and nano particles admixing [12-14]. Recent studies have shown that chemical doping may be effective and feasible approach for increasing critical current density of $MgB_2$ superconductors [15-18]. Therefore, it is necessary to study doping effect of suitable elements in $MgB_2$. This may open wide spread applications of $MgB_2$. In our recent study doping of La in $MgB_2$ has been found to increase critical current density due to $LaB_6$ nanoparticle inclusions in $MgB_2$ matrix [15].

Several researchers have reported the enhancement of $J_c$ by chemical doping. Dou et al., in a series of papers [19-21] have shown that SiC and carbon nanoparticle doping significantly improves $J_c$ & irreversibility field ($H_{irr}$). Mastsumoto et al [22] have reported enhancement of $J_c$ and $H_{irr}$ through $SiO_2$ and SiC doping.

In earlier studies Ag admixing has been reported to result in enhancement of critical current density in cuprate superconductors [23-25]. Recently studies on synthesis and superconducting characteristics of Ag admixed $MgB_2$ have been reported [26-28]. In these studies, authors have focused their investigation on low concentration of Ag admixed $MgB_2$ compound only and reported low critical current density e.g. Kumar et al. have observed intragrain $J_c$ value $\sim 1.5 \times 10^5$ A/cm$^2$ at 5K and in zero field [26]. However, a detailed investigation of role of Ag admixing in $MgB_2$ in rather wider compositional range leading to maximum enhancement of $J_c$ and $H_{irr}$ has not been carried out so far. In the present investigation, we have made an effort to find out optimum level and condition of admixing of Ag in $MgB_2$. Structural and microstructural characterization of Ag admixed $MgB_2$ superconductor has been done and the flux pinning capability of Ag has been explored. Based on the magnetization measurements, we have evaluated intragrain critical current density ($J_c$), behaviour of pinning force density ($F_P$) and upper critical field ($H_{c2}$) for Ag admixed samples.

**EXPERIMENTAL DETAILS**

Ag admixed $MgB_2$ bulk samples with nominal composition $MgB_2$–x at.% Ag ($0 \leq x \leq 30$ at.%) have been synthesized by solid-state reaction method at ambient pressure using high purity powders of Mg (99.9%), B (99%) and Ag (99.9%). The particles size of starting Mg, B and Ag powders are in the range of 30-40μm, ~ 5μm and 4-7μm respectively. These powders were fully mixed and cold pressed (3.5 tons/ inch$^2$) into small rectangular pellets (10 x 5 x 1 mm$^3$). Thereafter, the pellets were encapsulated in a Mg metal cover to take care of Mg loss and avoid the formation of MgO during the sintering process. The pellet configuration was wrapped in a Ta foil and sintered in flowing Ar atmosphere in a programmable tube type furnace at 900$^0$C for 2 hour. The pellets were cooled to room temperature at the rate of 5°C/ min. The encapsulating Mg cover was then removed and Ag added $MgB_2$ samples were retrieved for further studies. This encapsulation technique has been developed in our laboratory to synthesize $MgB_2$ superconductors [15].

All the samples in the present investigation were subjected to gross structural characterization by powder x-ray diffraction technique (XRD, PANalytical X` Pert Pro, CuK$_\alpha$) and microstructural characterization by transmission electron microscope (Philips EM-CM-12). The magnetization (M) measurements have been carried out at Tata Institute of Fundamental Research (Mumbai, India) over a temperature range of 5-40K employing a physical property measurement system (PPMS, Quantum Design) on fine ground powders of the as synthesized samples. Intragrain $J_c$ was calculated from the height 'ΔM' of the magnetization loop (M-H) using Bean's formula based on critical state model [29], It should be pointed out the Bean's formula leads to the optimum estimate of intragrain $J_c$ for superconductors having weakly coupled grains. However, this model will be appropriate for optimum estimation of $J_c$ in case of $MgB_2$ (where grains are strongly coupled) only when magnetization measurements are carried out on fine powder of the as synthesized sample. In the fine powder form, strong coupling is non-existent. Therefore, the $J_c$ can be estimated employing Bean's formula and using average size of the powder particles. It may be pointed out that fine ground particles usually correspond to agglomerates of nearly spherical shape (~ 5μm) covering only few grains (as estimated by scanning electron microscope SEM). Thus in the present investigation, we have used the average size of the powder particle (~ 5μm).

$$J_c = \frac{30 \Delta M}{\langle d \rangle}$$

Where 'ΔM' is the height of hysteresis loop in emu/cm$^3$ and <d> is the average particle size in cm (~5μm).

**RESULTS AND DISCUSSION**

The representative x-ray diffraction patterns of Ag admixed $MgB_2$ samples are shown in Fig.1 The XRD patterns reveal that all the samples are polycrystalline in nature and correspond to hexagonal structure of $MgB_2$ (a=b=3.08Å, c = 3.52Å). Any appreciable change in the lattice parameters of Ag admixed $MgB_2$ samples (using a computerized programme based on least square fitting method) has not been found within the experimental limit of 0.001Å  It may be noticed that Ag admixed $MgB_2$ samples have some additional peaks.  These additional peaks have been indexed to both Ag and MgAg for lower Ag concentration (Ag < 10 at.%).  However, for the higher Ag concentration (Ag ≥ 10 at.%), these additional peaks get explicable in terms of MgAg only. Unlike the case when Ag concentration is very small, for higher Ag concentration (Ag ≥ 10 at.%) the interfacial area of Mg/Ag will be large facilitating the interdifussion of Ag into Mg. This will lead to consumption of all Ag leading to the formation of MgAg phase only.

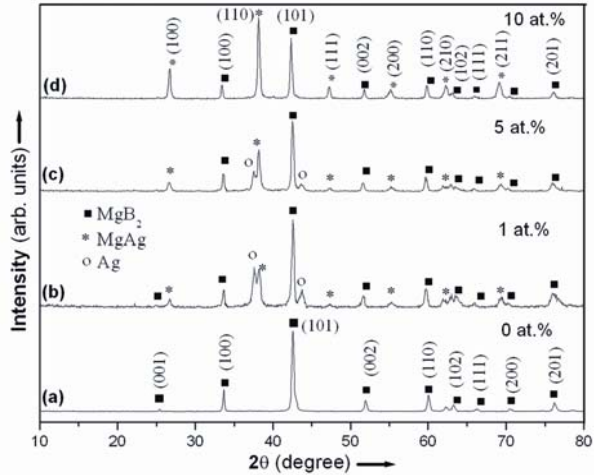 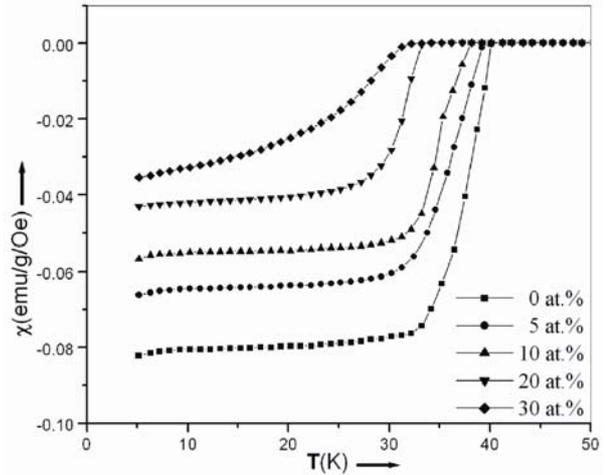

**Fig 1** *Representative powder XRD patterns of $MgB_2 - x$ at.% Ag (x=0, 1, 5 & 10 at.%).*

**Fig 2** *Temperature dependent dc magneti-susceptibility ($\chi$) behavior of $MgB_2 - x$ at.% Ag ($0 \leq x \leq 30$ at.%)*

The dc magnetic susceptibility ($\chi$) of $MgB_2$–x at.% Ag (with $0 \leq x \leq 30$ at.%) samples are shown in Fig.2 for 50 Oe field as a function of temperature. Based on this the transition temperature of $MgB_2$ admixed with different concentration of Ag can be taken to lie between 32 -40K. The decrease in $T_c$ with increasing concentration of Ag in the samples may be due to the presence of secondary phases (Ag, MgAg) in the sample.

The central aim of the present investigation is to explore the flux pinning properties and magnetic behaviour of Ag admixed $MgB_2$ samples and their possible correlation with microstructural features.  We therefore, first describe various microstructural features induced by different admixing concentrations of Ag in $MgB_2$.  Thereafter, evaluation of critical current density and behaviour of flux pinning force through magnetic measurements will be elucidated.  Finally correlation between intragrain $J_c$ and microstructural features will be described and discussed. The microstructural characterization has been carried out by transmission electron microscope in both imaging and diffraction mode of the as synthesized Ag admixed. $MgB_2$ samples with different Ag concentration. Fig.3(a) shows the representative transmission electron micrograph for $MgB_2$ compound. The selected area diffraction (SAD) pattern corresponding to the TEM micrograph is shown Fig.3(b), which reveals the hexagonal lattice pattern corresponding to $MgB_2$ compound.

With admixing of Ag in $MgB_2$ the dominant and specific microstructural feature is the occurrence of MgAg secondary nanoparticles, which are found to be invariably present. For example the presence of MgAg nanoparticles can be easily discernible from the representative TEM micrograph of $MgB_2$−10 at.% Ag compound [Fig.3(c)]. The density of the nanoparticles is higher at grain boundaries in comparison to within the $MgB_2$ grains. The average size of the nanoparticles inclusions has been found to be in the range of 5-20nm. Such a preferential presence of nanoparticles at grain boundaries (GBs) may be understood in term of interaction of grain boundary (GB) with the MgAg nanoparticles.  It may be pointed out that the GBs are disordered region in crystals. GBs therefore, represent high-energy configurations. Because of the Coulomb interaction between GB and the impurity atoms, GB tends to attract the impurity atoms in order to decrease its energy.  It should

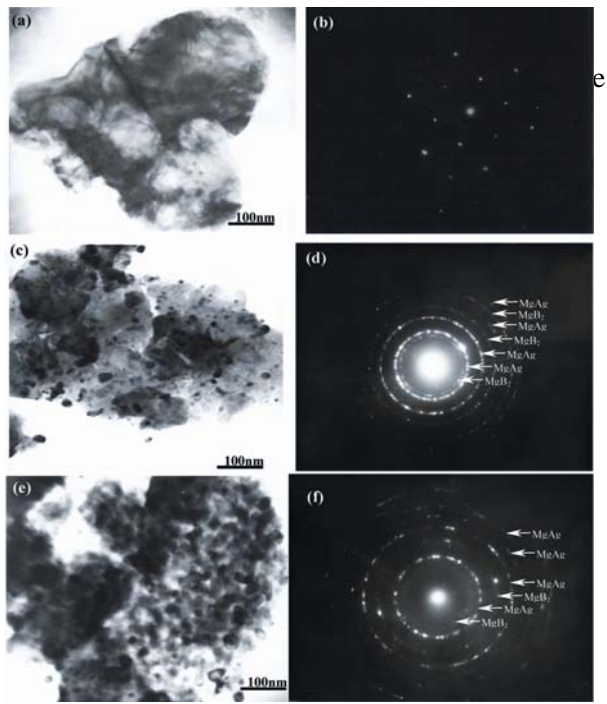
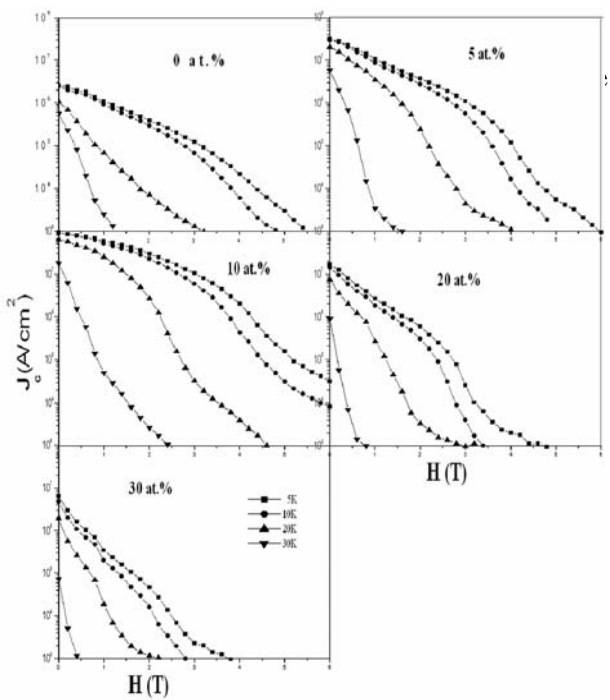

**Fig 3 (a)** *The representative TEM micrograph of pure $MgB_2$.* **(b)** *Selected area diffraction (SAD) pattern of hexagonal lattice corresponding to $MgB_2$ grain.*

**Fig 4** *Intragrain $J_c$ as a function of applied magnetic field for $MgB_2 - x$ at.% Ag ($0 \leq x \leq 30$ at.%) at 5, 10, 20 and 30K.*

*(c) Presence of MgAg nanoparticles discernible from the TEM micrograph of $MgB_2$ –10 at.% Ag admixed sample.(d) SAD pattern corresponding to TEM micrograph of fig.3 (a) shows spotty ring pattern which corresponds to $MgB_2$ and MgAg anoparticles (e) Representative TEM micrograph corresponding to $MgB_2$–30 at.% Ag admixed sample depicting significant precipitation of MgAg. (f) SAD pattern corresponding to TEM micrograph of fig. 3(e) has been indexed for $MgB_2$ and MgAg.*

While diffusion of Ag to the GB forming MgAg is easy because it will be quite difficult for Ag to diffuse into the grain because of large difference in the size of Mg and Ag. However, some of the Ag atoms would still diffuse into the grain because of Mg vacancies present in the grain and would form MgAg. Therefore, the chemical dopants have a higher probability to stay at the GB region. The SAD pattern corresponding to TEM micrograph [shown in Fig. 3(d)] reveals the spotty ring pattern. These diffraction rings, which correspond to $MgB_2$ and MgAg nanoparticles, depict the inclusion of nanoparticles in $MgB_2$.

The TEM micrograph for $MgB_2$–30 at.% Ag sample revealing the very high density of MgAg nanoparticles is discernible from the Fig. 3(e). It is interesting to note that distribution of MgAg nanoparticle for this sample is different from that of the $MgB_2$–10 at.% Ag sample. In this case density of nanoparticles is high within the $MgB_2$ grain. Such a feature may be due to high admixing concentration of Ag in $MgB_2$. The representative SAD pattern of $MgB_2$–30 at.% Ag [shown in Fig. 3(f)] have been indexed for $MgB_2$ and MgAg. Thus, it may be suggested that for lower concentration i.e. $x \leq 10$ at.%, there will be smaller concentration of MgAg in $MgB_2$ matrix in comparison to the grain boundaries. Upto this concentration of Ag (i.e. x~10 at.%), size and density of nanoparticles are suitable to act as flux pinning centres. When the Ag concentration is high i.e. $x > 10$ at.% significant precipitation/ segregation of AgMg results in the sample. Such a presence of high density of secondary phases leads to the suppression of superconductivity. It is interesting to notice that the definite formation of MgAg has been established through both XRD & SAD pattern (see Fig 1& 3). The exact reason of the formation of MgAg is not clear so far. However, it appears that the most feasible reason for the formation of MgAg is the reaction of excess Mg with Ag. It may be pointed out that in the present investigation, as outlined in experimental section, excess Mg was invariably taken to take care of Mg loss and also avoid the formation of MgO.

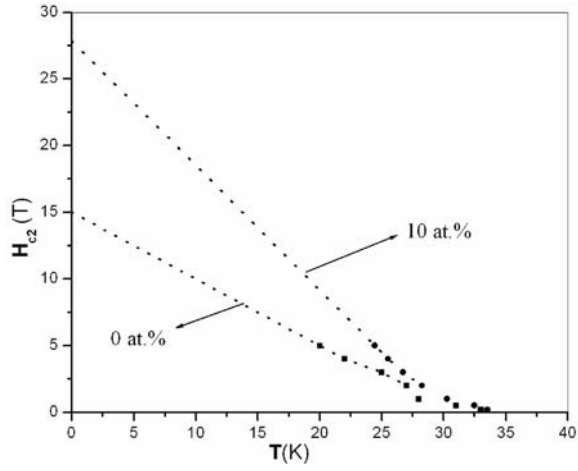 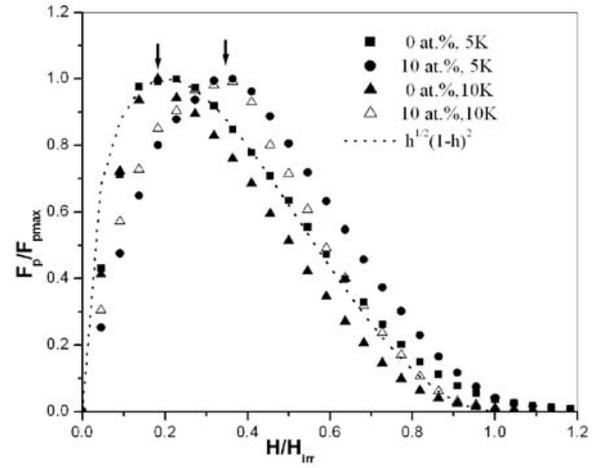

**Fig 5** *The extrapolated upper critical field as a function of temperature for $MgB_2$ and $MgB_2$ –10 at.% Ag admixed samples.*

**Fig 6** *Normalized pinning force $F_p/F_{pmax}$ as a function of reduced magnetic field for $MgB_2$ and $MgB_2$ –10 at.% Ag admixed samples at 5 & 10K. Shifting of peak position from Kramer plot towards higher magnetic field reveals the presence of extra pinning centres in Ag added $MgB_2$ sample.*

The magnetization measurements as a function of applied magnetic field (H) have been carried out at 5, 10, 20 and 30K, for each sample. The intragrain $J_c$ as a function of applied magnetic field $MgB_2$–x at.% Ag samples are shown in Fig. 4. It is clear from $J_c$ versus H curves that the intragrain $J_c$ of 10 at.% Ag sample attains the highest value among all the samples for all temperatures upto 30K and for the whole field region up to 5T. It appears from the present investigation that this sample contains optimum density of MgAg nanoparticles at GB as well as within the grain of $MgB_2$. For example at 5K, intragrain $J_c$ for 10 at.% Ag added sample is ~9.23 x $10^7$ A/cm$^2$ in self-field, ~ 5.82 x $10^7$ A/cm$^2$ at 1T,   ~ 4.24 x $10^6$ A/cm$^2$ at 3.6T and ~ 1.52 x $10^5$ A/cm$^2$ at 5T. The intragrain $J_c$ values for $MgB_2$ sample without Ag admixing are ~ 2.59 x $10^6$ A/cm$^2$, ~ 1.09 x $10^6$ A/cm$^2$, ~ 4.53 x $10^4$ A/cm$^2$ and 2.91 x $10^3$ A/cm$^2$ at 5 K in self field, 1T, 3.6T and 5T respectively. The above clearly shows that Ag admixing has resulted in enhancement of intragrain $J_c$ for all fields. This is consistent with microstructural characterization. As already outlined Ag admixing ≥10 at.% leads to the presence of discrete particles within the grain. The size of these particles is broadly compatible with the coherence length ~60Å. Similar variations of $J_c$ with magnetic fields were also observed for temperature 10, 20 and 30 K also. It should be pointed out that the value of $J_c$ found for 10 at.% in the present study is significantly higher than the values reported by earlier workers [26]. The values of $H_{c2}$, determined as the field at which M (H) first deviated from the background, as a function of temperature are shown in Fig. 5. The extrapolation of the curve gives the $H_{c2}$ value at 0K. The $H_{c2}$ value at 0K for $MgB_2$ sample without Ag is ~ 15T and for 10at.% Ag admixed $MgB_2$ sample is ~ 28T. These values of $H_{c2}$ are also close to the values obtained by Werthamer Helfand- Hohenberg model [30]

$$H_{c2}(0) = 0.7\, T_c \left( \frac{dH_{c2}}{dT} \right)$$

which yields 16T and 30T as $H_{c2}(0)$ values for pure $MgB_2$ and 10at.% Ag respectively.

The flux pinning mechanism associated with microstructural defects is often assessed by analyzing the slope of flux pinning force, $F_p(H)$, as a function of temperature. The normalized pinning force $F_p / F_{pmax}$ plotted against a reduced field h, (h = $H/H_{irr}$) typically overlap when a single pinning mechanism and pinning centre is dominant [31]. The irreversibility field ($H_{irr}$) has been estimated by extrapolating the $J_c^{1/2}H^{1/4}$ versus H curve to the horizontal axis. This technique (also called Krammer extrapolation) provides usually a very good estimate of $H_{irr}$[32]. In the present study the values of $H_{irr}$ at 10K are 4.7T and 5.8T for pure $MgB_2$ and optimally Ag admixed (10 at%) $MgB_2$ respectively. Such scaling behaviour is commonly observed in intermetallic low temperature superconductor (e.g. $Nb_3Sn$, NbTi)[33]. This pinning mechanism is governed by the shear modulus[32] and produces a bulk pinning force $F_p(H) = \mu_0 H.J_c(H)$ with the characteristics field dependence proportional to $h^{1/2}(1-h)^2$[34], where h is reduced magnetic field. In the present investigation there is a slight but clear shift of the peak of the pinning force towards higher field [see Fig. 6]. Such a shift of the peak of the pinning force is indication of the presence of additional pinning centres. Recently Cooley et al have shown that deviation from the usual flux shear behaviour is due to core pinning by small precipitation such as MgO nano precipitates in $MgB_2$ thin film[35]. Therefore, in the present study deviation of peak position from Kramer plot i.e. shifting of flux pinning force peak towards higher field, may be attributed to the nano inclusion of MgAg, which are expected to provide extra flux pinning force.

**CONCLUSION**

In conclusion, we have successfully synthesized Ag admixed $MgB_2$ sample at ambient pressure. In the present investigation exploration of microstructural features induced by admixing of Ag in $MgB_2$ compound and its correlation with intragrain $J_c$ have been carried out. The highest value of $J_c$ at 5K [~9.23 x $10^7$ $A/cm^2$ in self field, ~ 5.82 x $10^7$ $A/cm^2$, ~ 4.24 x $10^6$ $A/cm^2$ and ~ 1.52 x $10^6$ $A/cm^2$ at fields of 1T, 3.6T and 5T respectively] has been obtained for 10at.% sample. This enhancement of $J_c$ has been found to result due to optimum size and density of MgAg nanoparticles inclusions in $MgB_2$. The study of nature of flux pinning force shows a shift in peak positions from Kramer plot, which is due to core pinning by MgAg nanoparticles. This shifting of peak position leads us to conclude that the additional pinning centres are present in the samples in the form of MgAg nanoparticles

**ACKNOWLEDGEMENTS**

The authors are grateful to Prof. A.R. Verma, Prof. C.N.R. Rao, Prof. S.K. Joshi and Prof. A.K. Roychaudhary for fruitful discussion and suggestions. Financial supports from UGC, DST-UNANST and CSIR are gratefully acknowledged. One of the authors Rajiv Giri is thankful to CSIR New Delhi, Govt. of India for awarding SRF (Ext.) fellowship.